\documentclass[twocolumn,showpacs,preprintnumbers,amsmath,amssymb]{revtex4}
\usepackage{graphicx}
\usepackage{dcolumn}
\usepackage{bm}
\begin{document}
\preprint{APS/123-QED}
\title{Population of bound excited states in intermediate-energy fragmentation reactions}
\author{A.~Obertelli$^1$, A.~Gade$^1$, D.~Bazin$^1$, C.~M.~Campbell$^{1,2}$, J.~M.~Cook$^{1,2}$, P.~D.~Cottle$^3$, A.~D.~Davies$^{1,2}$, D.-C.~Dinca$^{1,2,\star}$, T.~Glasmacher$^{1,2}$, P.~G.~Hansen$^1$, T.~Hoagland$^1$, K.~W.~Kemper$^3$, J.-L.~Lecouey$^{1,\dagger}$, W.~F.~Mueller$^1$, R.~R.~Reynolds$^3$, B.~T.~Roeder$^3$, J.~R.~Terry$^{1,2}$, J. A.~Tostevin$^4$, K.~Yoneda$^{1,\ddagger}$, H.~Zwahlen$^{1,2}$}
\affiliation{$^1$National Superconducting Cyclotron Laboratory, Michigan State University, East Lansing, Michigan 48824, USA\\
$^2$Department of Physics and Astronomy, Michigan State University, East Lansing, Michigan 48824, USA\\
$^3$Department of Physics, Florida State University, Tallahassee, Florida 32306, USA\\
$^4$Department of Physics, School of Electronics and Physical Sciences, University of Surrey, Guildford, Surrey GU2 7XH, United Kingdom}
\altaffiliation{Present address: American Science \& Engineering, Inc., 829 Middlesex Turnpike, Billerica, Massachusetts 01821, USA\\
$^\dagger$Present address: Laboratoire de Physical Corpusculaire, 6 Boulevard du mar\'echal Juin, 14050 Caen Cedex, France \\
$^\ddagger$Present address: RIKEN, Hirosawa 2-1, Wako, Saitama 351-0198, Japan}

\date{\today}
\begin{abstract}
Fragmentation reactions with intermediate-energy heavy-ion beams exhibit a wide range of reaction mechanisms, ranging from direct reactions to statistical processes. We examine this transition by measuring the relative population of excited states in several $sd$-shell nuclei produced by fragmentation with the number of removed nucleons ranging from two to sixteen. The two-nucleon removal is consistent with a non-dissipative process whereas the removal of more than five nucleons appears to be mainly statistical. 
\end{abstract}
\maketitle
Beams of exotic nuclei provide a key tool to study nuclear structure far from stability. At intermediate energies ($\sim$100 MeV/nucleon), exotic beams can be produced via in-flight fragmentation, where the nucleus of interest is obtained by removing nucleons from an incident primary beam. During this reaction, the population of excited states depends on the number of nucleons removed. One-nucleon removal reactions from projectiles at intermediate energies have been shown to be mainly direct processes~\cite{1nucleon} with the resulting cross sections to different excited states sensitive to the quantum numbers of the removed nucleon. Two-nucleon knockout has recently been shown to be direct in certain energetically favorable cases~\cite{2nucleons,tostevin}. Fragmentation of heavy nuclei, on the other hand, can be described as a two-step process commonly called abrasion-ablation~\cite{aa}. During the abrasion stage the nucleons in the overlap volume of projectile and target are scraped off as the ions pass each other. In the ablation step the prefragment dissipates its excitation energy gained during the abrasion by particle emission~\cite{schmidt} and the different states of the residues are populated statistically with weights determined by the excitation energy of the prefragment. 
Few-nucleon removal reactions lie between these two limiting scenarios of purely direct and statistical production. Discrepancies between observed and predicted excited state populations \cite{morrissey1,nayak1,zhu} indicate the presence of non-statistical processes in few-nucleon removal reactions. In this article, we present new data on the relative population of bound excited states produced in fragmentation reactions and show that several-nucleon removal reactions at intermediate energies can be understood in terms of a smooth transition from direct to statistical processes.

In the experiment, we measured $\gamma$-rays in coincidence with fragmentation residues to determine the population of each of their bound states. Fragmentation residues were obtained from $^{34}$Ar, $^{26}$Si, $^{25}$Al and $^{24}$Mg secondary beams at $\sim$105 MeV/nucleon, which in turn were made by fragmentation of a 150 MeV/nucleon $^{36}$Ar primary beam from the Coupled Cyclotron Facility at the National Superconducting Cyclotron Laboratory (NSCL) at Michigan State University. The $^9$Be production target was located at the mid-acceptance target position of the A1900 fragment separator~\cite{a1900}. The exotic secondary beams impinged on a 376 mg/cm$^2$ $^9$Be secondary target located at the target position of the high-resolution, large-acceptance S800 spectrometer~\cite{s800}. The magnetic rigidity of the spectrometer was set to transmit the residues of interest, and prevent direct beam from entering the focal plane.  Each incident beam particle was characterized event-by-event in a time-of-flight measurement. Reaction residues were identified in the S800 focal plane through measurement of the energy loss in the ionization chamber of the spectrometer versus the time-of-flight measured between the object point and the focal plane. Gamma-rays from residues produced in bound excited states were detected in coincidence to determine the population of excited states. The secondary target was surrounded by the SeGA array~\cite{sega} composed of seventeen 32-fold segmented, high-purity Germanium detectors positioned at 20.8 cm from the center of the target. Seven detectors were placed at forward angles in a 37$^{\circ}$ ring with respect to the beam axis and ten detectors were located in a 90$^{\circ}$ ring around the target position. The photopeak efficiency of SeGA, in this configuration, is 2.5\% for a 1.3 MeV photon emitted from a moving source with a velocity of $v=0.35c$. The response of the detector was simulated with the GEANT code~\cite{geant} and the simulated detection efficiency validated with calibrated-source measurements. The population of excited states was determined from the $\gamma$-ray intensities relative to the number of reaction residues. The population to the ground-state was evaluated by subtraction. More than 90 \% of the momentum distribution was transmited for the studied two-nucleon removal reactions. For the other reaction channels, only part of the momentum distribution was within the acceptance of the S800, therefore no absolute cross sections could be measured. In the following, we assume that the momentum acceptance does not affect our conclusions.

\begin{figure}
\includegraphics[width=8.8cm]{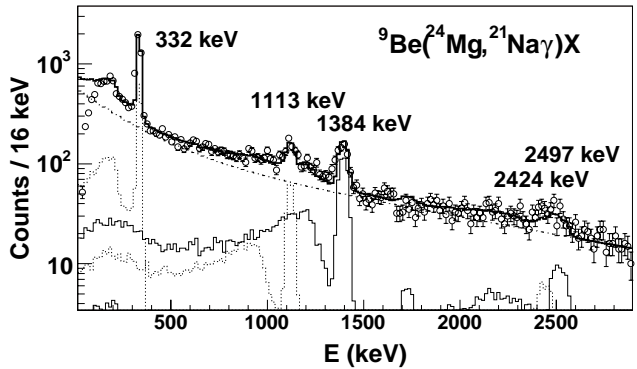}
\caption{$\gamma$-ray spectrum in coincidence with $^{21}$Na residues produced from $^{24}$Mg. The data points are shown with statistical error bars. The individual components of the simulated spectrum corresponding to individual $\gamma$-rays, the exponential background, and their sum (thick solid line) are presented.}
\label{fig0}
\end{figure}

The fragmentation residues analyzed here were chosen (i) to have a separation energy of less than 5 MeV to exclude potentially undetected high-energy feeding transitions and to thus ensure reliable relative population of individual excited states, and (ii) to be experimentally well-known and reproduced by shell-model calculations. With these criteria, we focused on the production of the $sd$-shell nuclei $^{21}$Na, $^{18}$Ne and $^{24}$Si, produced from different incoming beams. 

\begin{figure}
\includegraphics[width=8.8cm]{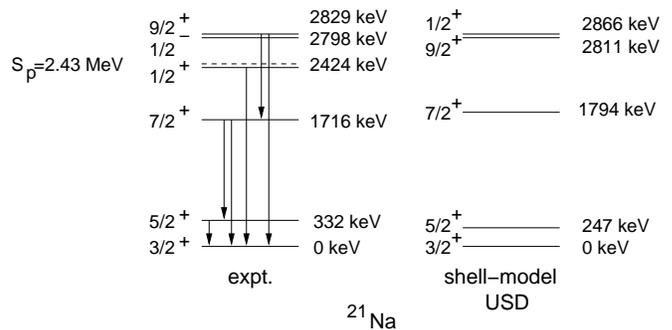}
\caption{Level-scheme of $^{21}$Na. Experimental data (left) are compared to shell-model calculations (right).}
\label{fig1}
\end{figure}

\begin{figure}
\includegraphics[width=8.8cm]{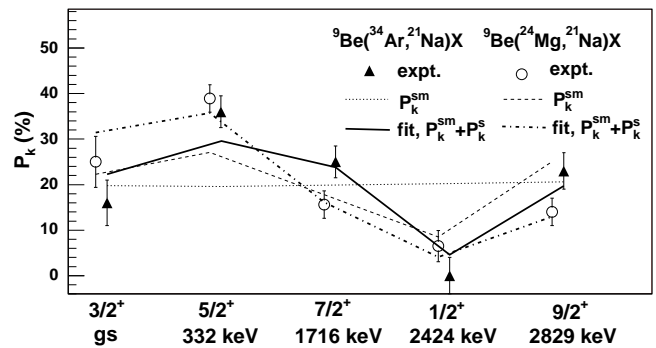}
\caption{Relative population of excited states in $^{21}$Na from $^{24}$Mg and $^{34}$Ar. For each reaction, the predictions P$^{sm}_k$ for a non-dissipative process and the fit with the population P$_k$=P$^{sm}_k$+P$^s_k$ are shown.}
\label{fig2}
\end{figure}

\begin{figure}
\includegraphics[width=8.8cm]{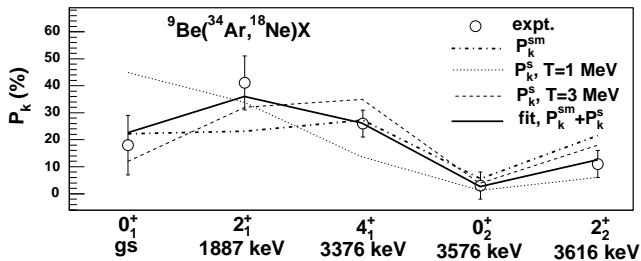}
\caption{Relative population of excited states of $^{18}$Ne from $^{34}$Ar. The predictions from non-dissipative population P$^{sm}_k$ (dotted-dashed line), statistical population P$^s_k$ with a temperature T=1 MeV (dotted line) and T=3 MeV (dashed line), and best fit from Eq.~\ref{fitotal} (solid line).}
\label{fig3}
\end{figure}

$^{21}$Na (S$_p$= 2.43 MeV) was produced from fragmentation of $^{24}$Mg, $^{25}$Al and $^{34}$Ar. Its $\gamma$-ray spectrum in coincidence with incoming $^{24}$Mg is shown in Fig~\ref{fig0}. $^{21}$Na presents three bound excited states and an unbound state that decays via electromagnetic transitions ($\Gamma_{\gamma} \gg \Gamma_p$~\cite{rolfs}). The level scheme of $^{21}$Na is compared to shell-model calculations performed with the code Oxbash~\cite{oxbash} and the USD interaction~\cite{usd} in Fig.~\ref{fig1}. 
The 7/2$^+$ excited state decays with 93(2)\%~\cite{firestone} to the 332 keV low-lying level via a 1384 keV transition, whereas the 2424 keV state decays directly to the ground-state. The proton-unbound 2829 keV level is known to decay via $\gamma$ emission with a 63(5)\% branching ratio to the 7/2$^+$ excited state and with 37(5)\% to the 332 keV level. The relative populations of these states in fragmentation of $^{24}$Mg and $^{34}$Ar are presented in Fig.~\ref{fig2}. A first comparison of the experimental data indicates that the fragmentation from $^{34}$Ar favors the high-spin states 7/2$^+$ and 9/2$^+$ relative to the fragmentation of $^{24}$Mg. The theoretical lines are discussed later.
 $^{18}$Ne was produced from $^{24}$Mg, $^{25}$Al, $^{26}$Si and $^{34}$Ar. It has four bound excited states below its proton separation energy of S$_p$=3922 keV. A 2$^{+}$ excited state at 1887 keV and a (0$^+$,2$^+$,4$^+$) triplet at $\sim$ 3500 keV.  The members of the triplet decay to the first 2$^{+}$ state. The relative population of states in $^{18}$Ne from the fragmentation of $^{34}$Ar is presented in Fig.~\ref{fig3}. The production of $^{24}$Si from $^{26}$Si was also studied as an interesting case of two-neutron removal. Two transitions were observed in $^{24}$Si (S$_p$= 3.3(4) keV) corresponding to two excited states previously reported in~\cite{schatz}. They were assigned as the first and second 2$^+$ states of $^{24}$Si at 1860(8) keV and 3410(14) keV respectively, based on good agreement with both shell-model calculations and the level scheme of the mirror nucleus $^{24}$Ne.\\ 
In the following, we present a simple approach to qualitatively reproduce our observations. The main idea is to consider the coexistence of both, statistical feeding of different states and a non-dissipative process which depends on the structure of the projectile and the residue.
In a first step, we consider a statistical population. Assuming a two-step fragmentation process with a thermalized prefragment and neglecting the effect of particle emission on the relative feeding of excited states in the residue, the statistical population P$^s_k$ of a residue's bound excited state $k$, with a spin j$_k$ and an excitation energy E$^{\star}_k$, is expected to be proportional to the Boltzmann factor weighted by the spin degeneracy~\cite{morrissey3}
\begin{equation}
\mbox{P}^s_k = N (2j_k+1) e^{-E^{\star}_k/T},
\label{stat}
\end{equation} 
where $N=1/(\sum_k \mbox{P}^s_k)$ is for normalization. In this parametrization, $T$ is not the usual temperature of the prefragment since P$^s_k$ is an estimation function for the population of the excited states of the \emph{residue}. $T$ represents  an average quantity including the effects of the temperature of the prefragment and the different decay channels of the prefragment to the residue. P$^s$ favors the population of high-spin and low-energy states. These statistical predictions are plotted for $^{18}$Ne in Fig.~\ref{fig3} for $T$=1~MeV and $T$=3~MeV. For the lower $T$, the population decays exponentially with the excitation energy E$^{\star}$. At $T$=3 MeV $\sim$ S$_p$=3.9 MeV, the population is driven by both the spin degeneracy of the excited states and the exponential decay with E$^{\star}$. Clearly, no value of $T$ can reproduce the observations within error bars.\\
\begin{figure}
\includegraphics[width=8.8cm]{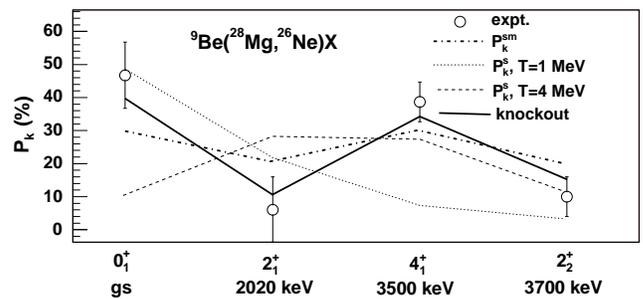}
\caption{Relative production of excited states of $^{26}$Ne from $^{28}$Mg~\cite{2nucleons}. Experimental data are compared to theoretical predictions, namely the population calculated from Eq. ~\ref{proba} (P$^{sm}_k$), a complete stripping calculation (knockout) and a statistical population (P$^s_k$) with T=1 MeV and T=3 MeV.}
\label{fig4}
\end{figure}
In a second step, we consider the population of excited states to be only due to non-dissipative processes, such as direct multi-nucleon removal or sequential stripping reactions. To estimate the relative populations of such mechanisms in a simple picture, we assume (i) that the relative production in these non-dissipative channels depends only on the projectile and residue shell-model wave functions and the single-particle removal cross sections, and (ii) that the cross sections for the removal of an $s$-wave and $d$-wave nucleon from the $sd$-shell are identical, as it is qualitatively justified by the single particle cross sections reported in~\cite{2nucleons}. We estimated an effective multi-nucleon removal strength between the projectile and the residue via shell-model wave functions. In a shell-model description, the wave-function $\vert \Phi_0 \rangle$ of the incoming particle with A nucleons is decomposed into its underlying independent-particle shell-model configurations $\vert \phi_i(A) \rangle$ with weight coefficients $w^0_i=\vert \langle \phi_i(A) \vert \Phi_0 \rangle \vert^2$. In the same way, the wave-function $\vert \Psi_k \rangle$ of the residue, produced in an excited state $k$ from a $\Delta$A-nucleon removal, is decomposed with the weight coefficients $\beta^k_j=\vert \langle \phi_j(A-\Delta A) \vert \Psi_k \rangle \vert^2$. The effective transition strengths $C_k$ between the incident particle and the residue in a state $k$, resulting from the removal of neutrons and protons, is taken as
\begin{equation}
C_k = \sum_i \sum_j w^0_i \beta^k_j c_{ij}, 
\end{equation}
where $c_{ij}$ is the number of ways one can remove $\Delta$A nucleons from $\vert \phi_i(A) \rangle$ to obtain $\vert \phi_j(A-\Delta A) \rangle$. $ c_{ij}$ is evaluated as
\begin{equation}
c_{ij}=\prod_{\theta} C(n^{\theta}_i(A),n^{\theta}_j(A-\Delta A)),
\end{equation}
where $C(n^{\theta}_i,n^{\theta}_j)$ is the number of combinations of $n^{\theta}_j$ nucleons choosen from a set of $n^{\theta}_i$ nucleons, and $n_i^{\theta}(A)$ is the number of nucleons of the configuration $\vert \phi_i(A) \rangle$ in the subshell $\theta \equiv (\ell,j,\tau)$, with $\ell$, $j$ and $\tau$ the angular momentum, total angular momentum and isospin of the considered subshell, respectively. $c_{ij}$ is non-zero only if the number of nucleons $n^{\theta}_i(A)$ in each occupied orbital $\theta$ of $\vert \phi_i(A) \rangle$ is greater or equal to the corresponding number of nucleons $n^{\theta}_j(A-\Delta A)$ in the same orbital of $\vert \phi_j(A-\Delta A) \rangle$. For simplicity, angular momentum couplings are neglected. The shell-model prediction P$^{sm}_k$ for the population of each bound state $k$ is then given by
\begin{equation}
\mbox{P}^{sm}_k=\frac{C_k}{\sum_k C_k}.
\label{proba}
\end{equation} 
 The calculation of P$^{sm}_k$ for a given system is parameter free. All calculations were performed within the $sd$ shell-model space with the shell-model code Oxbash~\cite{oxbash} and the USD interaction~\cite{usd}. The difference between P$^{sm}_k$ and a complete calculation is shown in the case of $^{26}$Ne produced by the two-proton knockout from $^{28}$Mg~\cite{2nucleons} in Fig.~\ref{fig4}. The stripping results, considering exact two-nucleon amplitudes and reaction cross sections, are taken from ref.~\cite{tostevin}. Eq.~\ref{proba} reproduces the trend of the complete calculation which is in good agreement with the data. This two-nucleon removal has been shown to be a direct process~\cite{2nucleons}. Thus a standard direct reaction mechanism is expected to predict the experimental feeding correctly. In this case, the statistical population (from Eq.~\ref{stat}) cannot reproduce the experimental behavior of the relative population within the experimental uncertainties, especially the strong population of both the ground state and the 4$^+$ state at E$^{\star}$=3500 keV. The dependence of P$^{sm}$ on the structure of the projectile for a given residue is clearly apparent in the case of $^{21}$Na produced from $^{24}$Mg and $^{34}$Ar (Fig.~\ref{fig2}). For a $^{34}$Ar projectile, the population function P$^{sm}$ is flat, whereas it shows large differences from state to state for incoming $^{24}$Mg.\\ 
Neither the statistical assumption nor the derived equation Eq.~\ref{proba} can reproduce the entire data set correctly. This is not surprising since the number of removed nucleons ranges from $\Delta$A=2, known to have a dominant direct part in the production mechanism, to $\Delta$A=16 expected to be well described by a two-step fragmentation statistical calculation. 

For this reason, we considered the coexistence of the two phenomena. The relative populations of each of the nine aforementioned systems were fitted with a function P$_k$ combining the statistical feeding P$^s_k$ with the non-dissipative population P$^{sm}_k$ in the following way
\begin{equation}
\mbox{P}_k = \alpha \mbox{P}^{sm}_k + (1-\alpha) \mbox{P}^s_k(T),
\label{fitotal}
\end{equation} 
where the weight of the single-particle component $\alpha$ and the temperature-like parameter $T$ of the statistical part were the only parameters. Both P$^s$ and P$^{sm}$ are normalized to unity so that $\sum_k P_k=1$. In adding the two probabilities, we neglected interferences between the two processes. For the two-nucleon removal, we considered the complete stripping reaction calculation for the direct single-particle function P$^{sm}$. The fitting was performed by minimizing a standard $\chi^2$ function with the following boundary conditions: $0 \leq \alpha \leq 1$ and $0 \leq T$. The results obtained are summarized in Table~\ref{summarize}. For each system, we report the $\chi^2$ values for the fit with Eq.~\ref{fitotal} (P$_k$), for the non-dissipative estimation P$^{sm}_k$, and for the fit with the statistical part P$^s_k$. The different $\chi^2$ values are not normalized to the number of degrees of freedom (depending on the number of parameters in the considered probability law) in order to offer an unbiased evaluation of the agreement of the data set and the theoretical assumption. The best fit with P$_k$ for each system reproduces the data well, as is illustrated by the good $\chi^2$ values.
\begin{table}
\caption{Comparison of data from this work and~\cite{2nucleons} with different assumptions (see text). The different reactions are classified with increasing number of removed nucleons $\Delta$A.}
\begin{ruledtabular}
\begin{tabular}{ccccccccc}
$\Delta$A&Res.&Proj.&Energy&$\alpha$&T&\multicolumn{3}{c}{$\chi^2$}\\
&&&(MeV/nucl.)&&(MeV)&P$_k$&P$^{sm}_k$&P$^{s}_k$\\
\hline
2&$^{26}$Ne&$^{28}$Mg&82&1.0(2)&-&0.9&0.9&9.2\\
2&$^{24}$Si&$^{26}$Si&109&0.8(2)&0.5(2)&0.0&2.3&19.5\\
3&$^{21}$Na&$^{24}$Mg&94&0.3(1)&1.1(1)&3.7&41.7&3.7\\
4&$^{21}$Na&$^{25}$Al&102&0.9(2)&0.3(1)&7.1&7.3&14.6\\
6&$^{18}$Ne&$^{24}$Mg&94&0.0(8)&1.2(3)&0.8&6.1&0.8\\
7&$^{18}$Ne&$^{25}$Al&102&0.0(9)&1.8(9)&0.3&1.5&0.3\\
8&$^{18}$Ne&$^{26}$Si&109&0.4(6)&1.0(5)&0.0&2.3&0.5\\
13&$^{21}$Na&$^{34}$Ar&110&0.0(1)&2.7(5)&7.0&50.5&7.0\\
16&$^{18}$Ne&$^{34}$Ar&110&0.0(8)&1.7(4)&0.2&2.7&0.2\\
\end{tabular}
\end{ruledtabular}
\label{summarize}
\end{table}
\begin{figure}
\includegraphics[width=8.8cm]{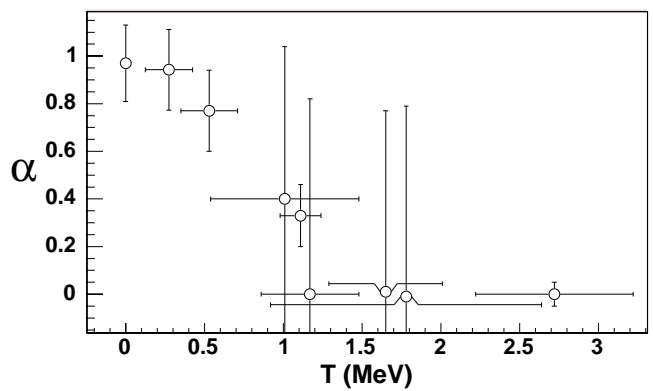}
\caption{Correlation between the weight $\alpha$ of direct process (see text) and the temperature-like parameter $T$ of the statistical part.}
\label{fig5}
\end{figure}  
For the residue $^{18}$Ne, the experimental uncertainties corresponding to the ground state and the first excited state are larger relative to other residues due to the decay scheme of $^{18}$Ne. Therefore, the fit is less sensitive, resulting in large error bars for the estimation of $\alpha$. The fit for the two-proton removal $^{9}$Be($^{28}$Mg,$^{26}$Ne)X gives $\alpha=0.97(11)$, in agreement with~\cite{tostevin}. For cases with the residue corresponding to $\Delta$A=6--16, the fit is consistent with $\alpha=0$, showing that for $\Delta$A $\gtrsim 5$ statistical population can account for the observations. Even if not strictly monotonic, the decrease of $\alpha$ is clearly correlated with the number of removed nucleons $\Delta$A. The weight $\alpha$ of the non-dissipative removal and the temperature-like parameter $T$ of the statistical process show a strong correlation as illustrated in Fig.~\ref{fig5}.  From the simple ingredients of the model, the observed correlation can be intuitively understood. At $T$=0 the population is expected to be non-statistical since P$^s$=0 except for the ground-state, whereas for increasing $T$, the stastistical feeding should be more dominant. This correlation validates the underlying intuitive assumptions of our approach.\\ 

We have fragmented $sd$-shell nuclei at intermediate energies and measured the relative feeding of excited states of the residues via $\gamma$-ray spectroscopy. Nine multi-nucleon removal reactions from $\Delta$A=2 to 16 have been studied. The two-nucleon removal is confirmed to be consistent with a non-dissipative, direct mechanism  whereas the removal of more than five nucleons is consistent with a purely statistical process. The results confirm that both a non-dissipative process and a statistical mechanism participate in multi-nucleon removal reactions when $2< \Delta \mbox{A} <6$ are being removed.\\

The authors acknowledge support from the U.S. National Science Foundation under Grants No. PHY-0110253 and No. PHY-0244453, and the United Kingdom Engineering and Physical Sciences Research Council (EPSRC) under Grant No. EP/D003628. We acknowledge valuable discussions with B.~A.~Brown and W.~G.~Lynch.

\end{document}